\documentclass[twocolumn,showpacs,preprintnumbers,amsmath,amssymb]{revtex4}

\usepackage{graphicx}
\usepackage{dcolumn}
\usepackage{bm}

\begin{document}

\preprint{APS/123-??}

\title{Charged antiparticle to particle ratios \\ near midrapidity in p+p collisions at $\sqrt{s_{_{NN}}}=200$ GeV}
\author{	
B.B.Back$^1$,
M.D.Baker$^2$,
M.Ballintijn$^4$,
D.S.Barton$^2$,
B.Becker$^2$,
R.R.Betts$^6$,
A.A.Bickley$^7$,
R.Bindel$^7$,
W.Busza$^4$,
A.Carroll$^2$,
M.P.Decowski$^4$,
E.Garc\'{\i}a$^6$,
T.Gburek$^3$,
N.George$^2$,
K.Gulbrandsen$^4$,
S.Gushue$^2$,
C.Halliwell$^6$,
J.Hamblen$^8$,
A.S.Harrington$^8$,
C.Henderson$^4$,
D.J.Hofman$^6$,
R.S.Hollis$^6$,
R.Ho\l y\'{n}ski$^3$,
B.Holzman$^2$,
A.Iordanova$^6$,
E.Johnson$^8$,
J.L.Kane$^4$,
N.Khan$^8$,
P.Kulinich$^4$,
C.M.Kuo$^5$,
J.W.Lee$^4$,
W.T.Lin$^5$,
S.Manly$^8$,
A.C.Mignerey$^7$,
R.Nouicer$^{2,6}$,
A.Olszewski$^3$,
R.Pak$^2$,
I.C.Park$^8$,
H.Pernegger$^4$,
C.Reed$^4$,
C.Roland$^4$,
G.Roland$^4$,
J.Sagerer$^6$,
P.Sarin$^4$,
I.Sedykh$^2$,
W.Skulski$^8$,
C.E.Smith$^6$,
P.Steinberg$^2$,
G.S.F.Stephans$^4$,
A.Sukhanov$^2$,
M.B.Tonjes$^7$,
A.Trzupek$^3$,
C.Vale$^4$,
G.J.van~Nieuwenhuizen$^4$,
R.Verdier$^4$,
G.I.Veres$^4$,
F.L.H.Wolfs$^8$,
B.Wosiek$^3$,
K.Wo\'{z}niak$^3$,
B.Wys\l ouch$^4$,
J.Zhang$^4$\\
\vspace{3mm}
\small
$^1$~Argonne National Laboratory, Argonne, IL 60439-4843, USA\\
$^2$~Brookhaven National Laboratory, Upton, NY 11973-5000, USA\\
$^3$~Institute of Nuclear Physics PAN, Krak\'{o}w, Poland\\
$^4$~Massachusetts Institute of Technology, Cambridge, MA 02139-4307, USA\\
$^5$~National Central University, Chung-Li, Taiwan\\
$^6$~University of Illinois at Chicago, Chicago, IL 60607-7059, USA\\
$^7$~University of Maryland, College Park, MD 20742, USA\\
$^8$~University of Rochester, Rochester, NY 14627, USA\\
}

\date{September 02, 2004}
\begin{abstract}\noindent
The ratios of the yields of primary charged antiparticles to particles have been obtained for pions, kaons, and protons near midrapidity for p+p collisions at $\sqrt{s_{_{NN}}} = 200$~GeV. Ratios of $\left<\pi^-/\pi^+\right>=1.000~\pm~0.012$ (stat.) $\pm~0.019$ (syst.), $\left<K^-/K^+\right>=0.93~\pm~0.05$ (stat.) $\pm~0.03$ (syst.), and $\left<\bar{p}/p\right>=0.85~\pm~0.04$ (stat.) $\pm~0.03$ (syst.) have been measured.  The reported values represent the ratio of the yields averaged over the rapidity range of $0.1<y_{\pi}<1.3$ and $0<y_{K,p}<0.8$, and for transverse momenta of $0.1<p_{T}^{\pi,K}<1.0$~GeV/c and $0.3<p_{T}^{p}<1.0$~GeV/c. Within the uncertainties, all three ratios are consistent with the values measured in d+Au collisions at the same energy. The data are compared to results from other collision systems and energies.
\end{abstract}


\maketitle

Relativistic heavy ion collisions allow for the study of strongly interacting matter in the high temperature and low baryochemical potential regime of the phase diagram of nuclear matter.
Smaller collision systems provide valuable reference information necessary to interpret the results obtained from more complicated heavy ion collisions.  
In this paper, the ratios of the yields of antiparticles to particles for primary charged pions, kaons, and protons emitted in p+p collisions at $\sqrt{s_{_{NN}}} = 200$~GeV, determined using data collected by the PHOBOS detector during the 2003 run of the Relativistic Heavy Ion Collider (RHIC), are presented. 

Comparisons of d+Au and Au+Au data suggest that the conditions created in the two collision systems differ \cite{dAhighpt}.   
Whether these different conditions influence the particle ratios was explored by comparing the values measured in d+Au and Au+Au collisions as a function of centrality \cite{phobosdAratios}.  
This study can be extended to lower multiplicity environments through the analysis of p+p collisions.  
To compare particle ratios measured in different collision systems the number of collisions experienced by each participating nucleon, $\nu$, is used as a measure of collision centrality.
This parameter is useful in the study of the $\bar{p}/p$ ratio which is sensitive to the interplay between the baryon number transport and antibaryon-baryon pair production processes.
In the p+p collision system each participating nucleon experiences only one collision.  
In the d+Au collision system, $\nu$ is defined as the number of collisions experienced by each participating nucleon from the deuteron, $\nu \equiv N_{coll}/N_{part}^{d}$, where $N_{coll}$ and $N_{part}^d$ are the number of binary collisions and the number of participants in the deuteron, respectively.
The d+Au particle ratios have been measured over a range of $\nu$ from 2 to 8 \cite{phobosdAratios}.  
Measurement of the antiparticle to particle ratios in the p+p collision system extends the range over which the ratios are known to the simplest possible nuclear system.  

The analysis reported here is nearly identical to that described in Ref. \cite{phobosdAratios} with two important exceptions. 
First, to increase the trigger efficiency, the primary event trigger was provided by two sets of 16 scintillator paddle counters located $-3.21$~m (PN) and $+3.21$~m (PP) from the nominal interaction point along the longitudinal ($z$) axis.
The paddle counters cover a pseudorapidity range of $3<|\eta|<4.5$.  
A triggered event required a coincidence between the paddle counters, PN and PP, within $\Delta t\leq$~$10$~ns.
For this data set, 29 million triggered events were collected.
Second, the offline event selection criteria were modified in accordance with the primary event trigger.  
A more restrictive timing requirement  was applied to the paddle coincidence ($\Delta t <5$~ns) to provide the vertexing algorithm \cite{phobosdAratios} with a restricted range within $75$~cm of the nominal vertex ($z=0$~cm).  
Events with a reconstructed vertex of $|z|<8$~cm were used to ensure that particles of both charges were properly reconstructed in the spectrometer for both magnet polarities.
For this range of vertices, the trigger samples $62\pm4$\% of the inelastic collision cross section. 
After requiring a valid vertex reconstruction, $48\pm5$\% of the inelastic cross section is available for analysis.

The tracking algorithm and particle identification cuts used in this analysis and consequently the kinematic acceptance are identical to those described previously \cite{phobosdAratios}.  
Pions, kaons and protons are measured over the rapidity ranges of $0.1<y_{\pi}<1.3$ and $0<y_{K,p}<0.8$.
As a function of transverse momentum, the acceptance for pions and kaons is $0.1<p_{T}^{\pi,K}<1.0$~GeV/c and is $0.3<p_{T}^{p}<1.0$~GeV/c for protons.
Table I provides a summary of the event and particle statistics in this data set.  

\begin{table}
\caption{Summary of analyzed event and particle counts.}
\begin{center}
\begin{tabular}{cccccccc}
 \hline
 \hline 
Polarity & Events & $\pi^{-}$ & $\pi^{+}$ & $K^{-}$ & $K^{+}$ & $\bar{p}$ & $p$\\ 
 \hline
 - 	& 2537906 & 5233 & 24478 & 287 & 899 & 506 & 1141 \\
 + 	& 1686214 & 16235 & 3500 & 551 & 190 & 585 & 396 \\
\hline
\hline 
\end{tabular}
\end{center}
\end{table}

Particles of a given charge bend in the same direction as do their oppositely charged counterparts in the reversed polarity magnetic field; hence, they have the same kinematic acceptance.
The average kinematic values ($\left< p_T \right>$, $\left< p_T^2 \right>$, $\left< y \right>$) for antiparticles and particles bending in the same direction are found to agree within $5\%$.  

The raw antiparticle to particle ratios are determined independently for each bending direction from the statistically weighted average of the ratios over subsets of the data, as described in Ref. \cite{phobos200}, thus eliminating the need to apply an acceptance correction.
Polarity-dependent systematic effects that are the same for the two bending directions, such as field strength dependence, are removed by averaging the ratios measured for the two bending directions \cite{phobos200}.
Polarity-dependent systematic effects that are different for each bending direction, such as the $z$-vertex distribution, are corrected for as described previously \cite{phobosdAratios}.  

The systematic errors in the ratios were determined by varying the criteria used for event selection, track selection, and particle identification. 
The magnitudes of the errors are of the order of $2$--$3\%$ and are similar to those reported for the d+Au data \cite{phobosdAratios}.  

The $\langle \bar{p} \rangle / \langle p \rangle$ ratio is corrected for the asymmetric absorption of antiprotons versus protons in the detector materials, contamination by secondary particles and feed-down from hyperon decay.  
The magnitudes of the correction factors are the same as reported in Ref. \cite{phobosdAratios} with the exception of the correction for secondary particles which was found to be $2.1\pm0.3\%$ (stat.) in the p+p data set and $1.6\pm0.3\%$ (stat.) in the d+Au data set.  
The total corrections to the pion and kaon data are small \cite{phobosdAratios} and are reflected in the final systematic errors of the ratios.

After applying all of the corrections, the antiparticle to particle ratios and associated statistical and systematic errors within the PHOBOS acceptance are:

$\left<\pi^-/\pi^+\right>$\hspace{0.12in}$= 1.000$\hspace{0.05in}$\pm~0.012$ (stat.)$~\pm~0.019$ (syst.),

$\left<K^-/K^+\right>  =   0.93$\hspace{0.12in}$\pm~0.05$\hspace{0.12in}(stat.)$~\pm~0.03$\hspace{0.12in}(syst.),

$~~\left<\bar{p}/p\right>$\hspace{0.22in}$=  0.85$\hspace{0.12in}$\pm~0.04$\hspace{0.12in}(stat.)$~\pm~0.03$\hspace{0.12in}(syst.).

\begin{figure}
\begin{center}
\includegraphics[width=3.3in]{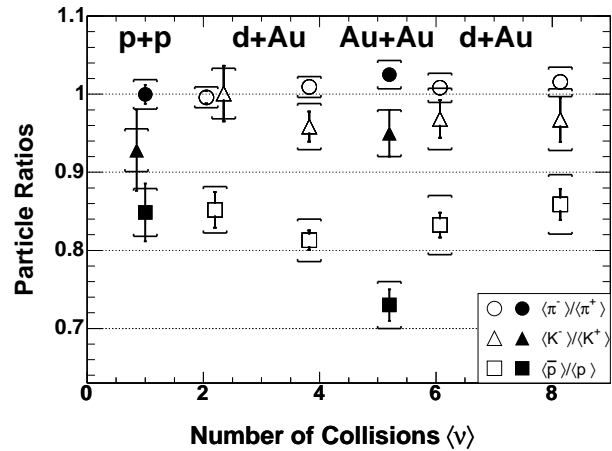}
\end{center}
\caption{Particle ratios as a function of centrality for each species for the p+p, d+Au and $12\%$ most central Au+Au collision systems at $\sqrt{s_{_{NN}}}=200$ GeV.  The brackets represent the total systematic error.} 
\label{AllRatios}
\end{figure}

Figure~\ref{AllRatios} shows the final particle ratios for the p+p, d+Au and $12\%$ most central Au+Au collisions at $\sqrt{s_{_{NN}}}=200$ GeV, plotted as a function of $\nu$ \cite{phobosdAratios, phobos200}.  
In the Au+Au collision system, $\nu$ is defined as $\nu \equiv \frac{N_{coll}}{N_{part}/2}$, where $N_{coll}$ and $N_{part}$ are the number of binary collisions and the total number of participants, respectively.
The errors shown in brackets represent the total systematic error.
For the d+Au data this quantity is calculated by adding the scale and point-to-point errors reported in Ref. \cite{phobosdAratios} in quadrature.
The d+Au and p+p data sets were both collected during the 2003 RHIC run while the Au+Au data are from the 2001 run.  
Within the errors, the ratios measured in the p+p and d+Au collision systems for all values of $\nu$ agree.  
However, the $\langle \bar{p} \rangle / \langle p \rangle$ ratio for Au+Au data is significantly lower than those observed in p+p and d+Au collisions.

\begin{figure}
\begin{center}
\includegraphics[width=3.3in]{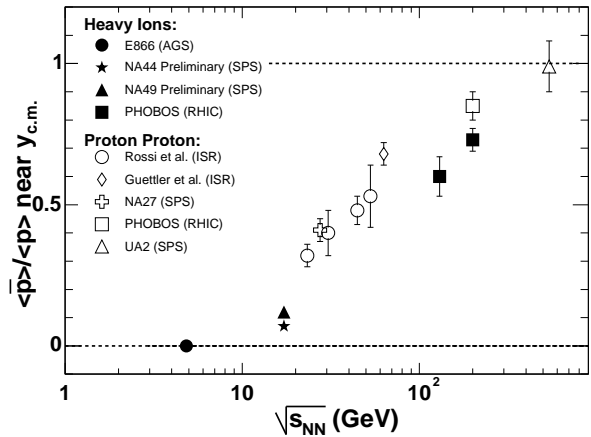}
\end{center}
\caption{Antiproton to proton ratio near midrapidity as a function of $\sqrt{s_{_{NN}}}$ for p+p collisions (open symbols) and central heavy ion collisions (filled symbols).  The error bars represent the statistical and systematic errors in the measurements added in quadrature.} 
\label{AllEnergies}
\end{figure}

The difference in the $\langle \bar{p} \rangle / \langle p \rangle$ ratio in p+p and Au+Au collisions can be investigated further by examining the evolution of the ratio as a function of collision energy.  
The $\langle \bar{p} \rangle / \langle p \rangle$ ratio near midrapidity as a function of center-of-mass energy for p+p collisions is shown in Fig.~\ref{AllEnergies} using open points \cite{compilation, guettler, ratioUA2, ratioNA27}.  
The data indicate a smooth evolution from low to high energy.  
Also shown in Fig.~\ref{AllEnergies} as filled symbols is the $\langle \bar{p} \rangle / \langle p \rangle$ ratio near midrapidity for central heavy ion collisions \cite{phobos200,phobos130,na44,na49,e866}. 
The $\langle \bar{p} \rangle / \langle p \rangle$ ratio in heavy ion collisions at RHIC energies has also been reported  at a variety of energies and centralities \cite{star130,brahms130, phenix130, brahms200, star200, phenixspectra}.  
In general, there is good agreement between the different measurements, but some small discrepancies remain to be resolved.
Figure~\ref{AllEnergies} presents the results from PHOBOS for 130 and 200 GeV Au+Au collisions \cite{phobos130,phobos200} which have similar acceptances as reported here for 200 GeV p+p collisions.  

The antiparticle to particle ratios have now been measured in both p+p and heavy ion collisions at the same energy using the same detector.
A suppression of the $\langle \bar{p} \rangle / \langle p \rangle$ ratio is observed in the heavy ion data relative to p+p over approximately an order of magnitude in the collision energy.  
The evolution of the ratio as a function of center-of-mass energy seems to be similar in shape for the two systems, with the heavy ion data offset by a factor of about two in collision energy.
The results reported in this paper will provide input to theoretical comparisons of the relative degree of baryon transport and antibaryon-baryon pair formation observed in p+p and heavy ion collisions and provide further information concerning the different dynamics that influence the evolution of each system.

This work was partially supported by U.S. DOE grants DE-AC02-98CH10886, DE-FG02-93ER40802, DE-FC02-94ER40818, DE-FG02-94ER40865, DE-FG02-99ER41099, and W-31-109-ENG-38, by U.S. NSF grants 9603486, 0072204, and 0245011, by Polish KBN grant 1-PO3B-062-27(2004-2007), and by NSC of Taiwan Contract NSC 89-2112-M-008-024.

\end{document}